\documentclass[runningheads]{llncs}
\let\llncssubparagraph\subparagraph
\let\subparagraph\paragraph
\usepackage[compact]{titlesec}
\let\subparagraph\llncssubparagraph

\usepackage[T1]{fontenc}
\usepackage{graphicx}
\usepackage{glossaries}

\usepackage{hyperref}
\usepackage{color}

\usepackage{cleveref}

\usepackage{enumitem}

\usepackage{todonotes}

\usepackage{titlesec}

\titlespacing*{\paragraph}{0pt}{1.5ex plus 0.5ex minus 0.5ex}{0.5em}
\titlespacing*{\subsection}{0pt}{2.5ex plus 0.5ex minus 0.2ex}{0.5ex minus 0.2ex}

\usepackage{lipsum}

\newacronym{faang}{FAANG}{Facebook, Amazon, Apple, Netflix, and Google}

\newacronym{it}{IT}{Information Technology}
\newacronym{xml}{XML}{eXtensible Markup Language}
\newacronym{api}{API}{Application Programming Interface}

\newacronym{mdse}{MDSE}{Model Driven Software Engineering}

\newacronym{is}{IS}{Information System}
\newacronym{bpm}{BPM}{Business Process Management}
\newacronym{bpm-lifecycle}{BPM-lifecycle}{Business Process Management lifecycle}
\newacronym{bpms}{BPMS}{Business Process Management System}
\newacronym{rpa}{RPA}{Robotics Process Automation}

\newacronym{ai}{AI}{Artificial Intelligence}
\newacronym{ml}{ML}{Machine Learning}
\newacronym{kg}{KG}{Knowledge Graph}
\newacronym{pca}{PCA}{principal component analysis}
\newacronym{nn}{NN}{Neural Networks}
\newacronym{dl}{DL}{Deep Learning}
\newacronym{llm}{LLM}{Large Language Model}
\newacronym{gpt}{GPT}{Generative Pre-trained Transformer}
\newacronym{opt}{OPT}{Open Pre-trained Transformer}
\newacronym{bloom}{BLOOM}{BigScience Large Open-science Open-access Multilingual Language Model}
\newacronym{nlp}{NLP}{Natural Language Processing}
\newacronym{rlhf}{RLHF}{Reinforcement Learning from Human Feedback}

\newcommand{\repeatthanks}{\textsuperscript{\thefootnote}}

\newcommand\blfootnote[1]{%
  \begingroup
  \renewcommand\thefootnote{}\footnote{#1}%
  \addtocounter{footnote}{-1}%
  \endgroup
}

\begin{document}
\title{Large Language Models for Business Process Management: Opportunities and Challenges\thanks{This research received funding from the \href{https://www.teamingai-project.eu/}{Teaming.AI} project, which is part of the European Union's \href{https://ec.europa.eu/programmes/horizon2020/en/home}{Horizon 2020} research and innovation program under grant agreement \href{https://cordis.europa.eu/project/id/957402/de}{No 957402}. The research by Jan Mendling was supported by the Einstein Foundation Berlin under grant EPP-2019-524 and by the German Federal Ministry of Education and Research under grant 16DII133.}
}
\titlerunning{\acrshort{llm} for \acrshort{bpm}: Opportunities and Challenges}
% If the paper title is too long for the running head, you can set
% an abbreviated paper title here
%
\author{
Maxim Vidgof\thanks{Equal contribution}\inst{1}\orcidID{0000-0003-2394-2247}
\and
Stefan Bachhofner\repeatthanks\inst{1}\orcidID{0000-0001-7785-2090}
\and 
Jan Mendling\inst{1,2,3}\orcidID{0000-0002-7260-524X}%
}
\authorrunning{M. Vidgof et al.}
% First names are abbreviated in the running head.
% If there are more than two authors, 'et al.' is used.
%
\institute{
Vienna University of Economics and Business,
Welthandelsplatz 1, 1020 Vienna, Austria
\email{\{first\_name.last\_name\}@wu.ac.at}
\and 
Humboldt-Universität zu Berlin, Unter den Linden 6, 10099 Berlin, Germany \\
\email{\{jan.mendling\}@hu-berlin.de} \and
Weizenbaum Institute, Hardenbergstraße 32, 10623 Berlin, Germany
}
\maketitle              
% typeset the header of the contribution
%

\begin{abstract}
Large language models are deep learning models with a large number of parameters.
The models made noticeable progress on a large number of tasks, and as a consequence 
% making them useful tools for a range of applications.
allowing them to serve as valuable and versatile tools for a diverse range of applications.
%They can for example be used to summarize a process description or an academic paper, support process modelers in explaining process models, and even propose research questions for a field or objections to an argument.
Their capabilities also offer opportunities for business process management, however, these opportunities have not yet been systematically investigated.  %practitioners, for example on the usage of these language models within a project. And for researchers on how they conduct research, for example on how research papers are developed.
In this paper, we address this research problem by foregrounding various management tasks of the BPM lifecycle. 
%systematically discuss the promises of these large language models for business process management.
%Building on this discussion, we use the business process management lifecycle to present possible applications of large language models.
We investigate six research directions highlighting problems that need to be addressed when using large language models,
% One of these research directions is 
including usage guidelines for practitioners.

% The abstract should briefly summarize the contents of the paper in
% 150--250 words.
\keywords{
    Natural language processing  \and 
    Large language models \and 
    Generative Pre-Trained Transformer \and
    Deep learning \and
    Research Challenges
}
\end{abstract}

\section{Introduction}
\blfootnote{\emph{Preprint submitted to arXiv}}
%
% Attention
%
Recent releases of applications building on \gls{llm} have been quickly adopted by large circle of users. ChatGPT stands out with reaching 100 million users in 2 months~\cite{Teubner2023}. The key factor explaining this fast uptake is their general applicability making them a general-purpose technology. Also many tasks in research can be approached with \gls{llm} applications, include finding peer reviewers, evaluating manuscripts and grants, improving prose in manuscripts, and summarizing texts~\cite{van2022language}.
%\gls{llm} can be used for many more applications within science, making them a versatile tool.
For this reason, some argue that \gls{llm} -- especially conversational \gls{llm} -- are a \emph{''game-changer for science''}~\cite{Van_Dis2023-ys}.

%
% Interest
%
Much of the current discussion of applications like ChatGPT is concerned with the question how good it works now and in the future. We believe that this question needs to be approached with a clearly defined task in mind. Starting with a task focus will move the discussion away from funny or disturbing errors and biases~\cite{Teubner2023} towards how the collaboration between human experts and \gls{llm} applications can be organized. Furthermore, this bears the chance to learn about specific categories of failures, which eventually will help to refine the technology in a systematic way.

%
% Desire
%
In this paper, we address the research challenge of how \gls{llm} applications can be integrated at different stages of business process management. To this end, we refer to the BPM lifecycle~\cite{dumas2018fundamentals} and its various management tasks~\cite{malinova2018identifying}. Our research approach is exploratory in a sense that we developed strategies of how \gls{llm} applications can be integrated in specific BPM tasks. We observe various promising usage scenarios and identify challenges for future research. 

%
% Action
%
The paper is structured as follows. 
\Cref{sec:background} discusses the essential concepts of \gls{dl} and \gls{llm} in relation to \gls{bpm} practises.
%\todo{fix}\Cref{sec:related:work} we relate this paper to previous research on novel problems caused by and related to \gls{ai} technologies, including \glspl{kg}, \gls{ml}, and \gls{nlp}.
In \Cref{sec:large:language:models:and:the:bpm:lifecycle} we identify and discuss \gls{llm} applications within \gls{bpm} and along the different \gls{bpm} lifecycle phases.
Based on these applications, \Cref{sec:research:directions} describes six core research directions ranging from how \gls{llm} change the dynamics and execution \gls{bpm} projects, to data sets, and benchmarks specific to \gls{bpm}.
\Cref{sec:discussion} identifies challenges when using \gls{llm}. Furthermore, we provide an outlook on how \gls{llm} might evolve in the future. %, and, finally, argue for the importance of combining different technologies well, which we believe will be crucial in fully exploiting the potential gains of \gls{llm}.

%In \Cref{sec:background}, we present the most important concepts from \gls{nlp}, \gls{dl}, \gls{llm}, and discuss how \gls{llm} pose new problems for \gls{bpm} in practise and research.
%In \todo{fix}\Cref{sec:related:work} we relate this paper to previous research on novel problems caused by and related to \gls{ai} technologies, including \glspl{kg}, \gls{ml}, and \gls{nlp}.
%In \Cref{sec:large:language:models:and:the:bpm:lifecycle} we identify and discuss \gls{llm} applications within \gls{bpm} and organize these applications with the \gls{bpm}-lifecycle.
%Based on these applications, we identify six core research directions ranging from how \gls{llm} change the dynamic and execution \gls{bpm} projects to data sets and benchmarks specific to \gls{bpm} in \Cref{sec:research:directions}.
%In \Cref{sec:discussion}, we list challenges when using \gls{llm} and future work, provide an outlook on how \gls{llm} might evolve in the future, and, finally, argue for the importance of combining different technologies well, which we believe will be crucial in fully exploiting the potential gains of \gls{llm}.

\section{Background}
\label{sec:background}
The advent of \gls{llm} applications paves the way towards a plethora of new BPM-related applications. So far, BPM has adopted natural language processing~\cite{van2018challenges}, artificial intelligence~\cite{dumas2023ai}, and knowledge graphs~\cite{10.1007/978-3-031-23515-3_7} to support various application scenarios. In this section, we discuss the foundations of %\gls{nlp} (\Cref{subsec:natural:language:processing}), 
\gls{dl} (\Cref{subsec:deep:learning}) and \glspl{llm} (\Cref{subsec:large:language:models}). In this way, we aim to clarify their specific capabilities. % and how these developed.

\subsection{Deep learning}
\label{subsec:deep:learning}
Recent \gls{llm} applications build on machine learning and deep learning models, such as recurrent neural networks (RNNs) and transformer networks. 
%\paragraph{Machine learning}
\gls{ml} studies algorithms that are \emph{''capable of learning to improve their performance of a task on the basis of their own previous experience''}~\cite{mjolsness2001machine}. 
%Prominent examples of such algorithms are \gls{pca}, decision trees, random forests, and \gls{nn}. The algorithms developed, and subsequently studied, by \gls{ml} have progressed immensely in the last two decades~\cite{jordan2015machine}. As of today, they are widely used in academia and commercial products, from computer vision, speech recognition, natural language processing, and robot control~\cite{jordan2015machine}. One major dimension on how to categorize \gls{ml} algorithms is based on the information that is available in the training data, more specifically, the feedback signal.
In essence, \gls{ml} techniques use either supervised learning, unsupervised learning, and reinforcement learning as a paradigm. Several of them are relevant for \gls{llm}.

%\paragraph{Supervised learning, Unsupervised learning, and Reinforcement learning}
In \emph{supervised learning}, the \gls{ml} algorithm receives as an input a collection of pairs, where one pair consists of features representing a concept, along with a label.
Importantly, this label is task specific and encodes what the algorithm should learn about the concepts.
%Examples of concepts are images and texts.
%A given image may be represented by a red, green, and blue feature tensor, or the same image can be represented by a gray scale feature tensor.
Such labels can be, for instance, \emph{spam} and \emph{no spam} for a spam classifier, or bounding boxes with annotations for an image.
%As is evident from the labels, the key in supervised learning is that we want the algorithm to perform well on a well defined task.
%In the \gls{ml} literature this is referred to as task specific.
%With reference to these labels, \gls{ml} differentiates between two types of tasks - on one hand classifications (such as predicting spam or no spam) or regression (such as bounding box regression).
%Finally, 
There are two cases of supervised learning that are relevant for \gls{llm}: \emph{few-shot} and \emph{zero-shot} learning.
\emph{Few-shot} learning is when a \gls{ml} algorithm adapts to a new situation with little amount of labelled data, and \emph{zero-shot} learning is when the algorithm can do this with no labelled data at all.
%Note here that the definition does not imply any change in the parameters of the model.
For example, a language model can be provided with a few input-output pairs, and the model can inverse the mapping function without any parameter changes.
In \emph{unsupervised learning}, the algorithm only receives a feature tensor of a concept as an input and the desired output is unknown.
The algorithm then finds structural properties of the concepts present in the feature tensor.
%Two big problems this learning paradigm studies are dimensionality reduction and clustering.
A typical application is dimensionality reduction, for instance using auto-encoders. %, the assumption is that the feature tensor can be reduced to a much lower dimensional representation - and the goal is to find this representation. Popular methods for this task are \gls{pca} and auto-encoders. In clustering, the assumption is the presence of unknown groups which form a partition (called clusters) in the feature tensor. The key here is that the clusters are unknown - and hence we can not provide any labels. Prominent examples of clustering algorithms are k-means and hierarchical clustering.
In \emph{reinforcement learning}, the algorithm receives a feature tensor of a concept as an input for which an output is produced, which is then evaluated through rewards.
The algorithm then uses this feedback to improve its parameters.
%This type of learning is used for tasks where the goal is to take actions in an environment to maximize a reward signal.
%It is commonly used in robotics, gaming and control systems, among other areas.
%The algorithm learns through trial and error, receiving positive or negative rewards for its actions, and adjusts its decision-making process accordingly.
%In this paradigm, the algorithm acts as an agent that interacts with its environment and learns how to optimize its behavior based on the rewards it receives.
ChatGPT uses a form of reinforcement learning known as deep reinforcement learning to improve its language generation capabilities, in particular, \emph{''Learning to summarize from human feedback''}~\cite{NEURIPS2020_1f89885d}.
ChatGPT is fine-tuned using a reward signal that assesses the quality of its generated responses, with the goal of maximizing the reward signal over time. The model's ability to learn from the reward signal allows it to generate increasingly relevant and coherent responses.

Deep learning (\gls{dl}) is a \gls{ml} method based on \gls{nn}. 
In general, they are \glspl{nn} with many layers stacked on top each other, which enables them to learn multiple layers of representations~\cite{lecun2015deep}.
Importantly, these representations can be learned without supervision.%, which is one key success factor.
%Shallow \gls{nn}, on the other hand, have 
Networks with only one hidden layer are called shallow.% and are therefore not able to learn a hierarchy of representations. 
Deep networks are able to handle more complex problems compared to shallow networks. 
Combined with the availability of large amounts of data, improvements on how to speed up the optimization, and powerful computing resources, enables them to be trained effectively.
%One of the key challenges in deep learning is to avoid overfitting, which is when the network memorizes the training data, instead of learning a general relationship between inputs and outputs. 
%This can be prevented through various techniques such as regularization and early stopping.
In the context of natural language processing, deep learning has been particularly effective in tasks such as machine translation, sentiment analysis, and named entity recognition. 
The ability of deep learning to learn multiple layers of representations from input data has proven to be particularly powerful for these tasks. 
This is because natural language processing involves dealing with sequences of words and characters, and the relationships between these sequences are often complex and multi-layered.
%\paragraph{Deep learning for natural language processing}
%\gls{dl} has proven to be effective in a wide range of \gls{nlp} tasks, including sentiment analysis, text classification, machine translation, and text generation. The power of deep learning for \gls{nlp} lies in its ability to automatically learn complex representations of the input data without the need for manual feature engineering. 
The use of large amounts of labeled training data and powerful computational resources has enabled deep learning models to achieve state-of-the-art results in many \gls{nlp} tasks. 
For example, the transformer architecture, introduced in the paper \emph{''Attention is All You Need''} by Vaswani et al.~\cite{vaswani2017attention} has become the standard architecture for many NLP tasks, including language translation and language modeling.

In recent years, BPM research has integrated the capabilities of deep learning to a large extent for process prediction. For an overview, see~\cite{neu2022systematic}. There are also recent applications for automatic process discovery~\cite{sommers2023supervised}, for generating process models from hand-written sketches~\cite{schafer2022sketch2process}, and for anomaly detection~\cite{nolle2020deepalign}.

\subsection{Large language models}
\label{subsec:large:language:models}

%
% Add figure showing the growth of number of parameters and training data set size
% https://research.aimultiple.com/gpt/
%
%
%\paragraph{Large language models}
\gls{llm} are \gls{dl} models trained on vast amounts of text data to perform various natural language processing tasks. 
These models, which typically range from hundreds of millions to billions of parameters, are designed to capture the complexities and nuances of human language. 
The largest models, such as \acrshort{gpt}-1 and \acrshort{gpt}-3, are capable of generating human-like text, answering questions, translating languages, and computer code.
The training process of these models involves processing massive amounts of text data, which is used to learn patterns and relationships between words and phrases. 
These models then use this information to predict the likelihood of a given token, or sequence of tokens, in a specific context. 
This allows them to generate coherent and contextually relevant text or perform other language-related tasks.
The rise of large language models has resulted in significant advancements in the field of NLP, and they are widely used in various applications, including chatbots, virtual assistants, and text generation systems.
One of their strengths is their ability to perform \emph{few-shot} and \emph{zero-shot} learning with prompt-based learning~\cite{liu2023pre}.

%
% Look up calculation of parameters
%
%\paragraph{GPT-1}
In 2018, Radford et al. introduced GPT-1 (also sometimes called simply GPT) in their paper on \emph{''Improving language understanding by generative pre-training''}
%~\footnote{See the \href{https://github.com/openai/finetune-transformer-lm}{GitHub repository of GPT-1} for more information.}~
\cite{radford2018improving}.
\gls{gpt}-1 refers to the largest model the authors have trained (110 million parameters).
In the paper, the authors studied the ability of transformer networks trained in two phases for language understanding.
In the first phase, %the authors trained a transformer network unsupervised - in particular 
they trained a transformer network to predict the next token given a set of tokens that appeared before (also called unsupervised pre-training, generative pre-training, or in statistics auto-regressive).
In the second phase, the transformer networks was fine tuned on tasks with supervised learning (also called discriminative fine-tuning).
In summary, their major finding is that combining task agnostic unsupervised learning in the first phase, then using this model in a second phase with supervised learning for fine tuning on tasks can lead to performance gains - from 1.5\% on textual entailment to 8.9\% on commonsense reasoning.

%\paragraph{GPT-2}
In 2019, Radford et al. introduced GPT-2 in their paper \emph{''Language Models are Unsupervised Multitask Learners''}%~\footnote{See the \href{https://github.com/openai/gpt-2}{GitHub repository of GPT-2} for more information.}~
\cite{radford2019language}.
Again, \gls{gpt}-2 refers to the largest model they have trained. GPT-2 is hence a scaled up version of GPT-1 in model size (1.5 billion parameters), and also in training data size.
In particular, GPT-2 has roughly more than ten times the number of parameters than GPT-1, and is trained on roughly more than ten times the amount of training data.
They report two major findings.
First, the unsupervised GPT-2 can outperform language models that are trained on task specific data set, without these data sets being in the training data set of GPT-2.
Second, GPT-2 seems to learn tasks (for example question answering) from unlabeled text data.
In both cases, however, the performance did not reach the state-of-the-art.
In summary, their major finding is that \glspl{llm} can learn tasks without the need to train them on these tasks, given that they have sufficient unlabeled training data.

%\paragraph{GPT-3}
In 2020, Brown et al. introduced \gls{gpt}-3 with the paper \emph{''Language Models are Few-Shot Learners''}%~\footnote{See the \href{https://github.com/openai/gpt-3}{GitHub repository of GPT-3} for more information}~
\cite{NEURIPS2020_1457c0d6}.
Unlike the above two cases, \gls{gpt}-3 refers to all the models the authors have trained, i.e. it refers to a family of models. 
The largest model the authors have trained is \gls{gpt}-3 175B, a model with 175 billion parameters.
In their paper, the authors showed that language models like GPT-3 can learn tasks with only a few examples, hence the title includes \emph{''few-shot learners''}. 
The authors demonstrated this ability by fine-tuning GPT-3 on various tasks, including question answering and language translation, using only a small number of examples.

%\paragraph{GPT-4}
In 2023, OpenAI introduced \gls{gpt}-4~\cite{OpenAI:GPT4}.
Contrary to previous versions of \gls{gpt}, this version is a multimodal  model as it can process text and images as an input to produce text.
This model is a major step forward as it improves on numerous benchmarks; however, it suffers from reliability issues, a limited context window, and inability to learn from experience like previous \gls{gpt} models.
%Among these issues, the authors list \emph{''it is not fully reliable ..., has a limited context window, and does not learn from experience''}.
%For this reason, the authors also provide a comprehensive system card where they describe these issues in greater detail.
This release, however, diverges from previous \gls{gpt} models as OpenAI is secretive about \emph{''the architecture (including model size), hardware, training compute, dataset construction, training method, or similar''}.
We only know about the model that it is a transformer style-model, pre-trained on predicting the next token on publicly available and not disclosed licensed data, and then fine-tuned with \gls{rlhf}.
Notwithstanding this departure, the authors include in their report findings on predicting model scalability.
They in particular report on predicting the loss as a function of compute, and the mean log pass rate (a measure on how many code sample pass a unit test) as a function of compute given a training methodology.
In both cases, they find that they could predict the respective measure with high accuracy based on data generated with significantly less compute (1.000 to 10.000 less).
They also find the inverse scaling price for a task, meaning that the performance on a task first decreases as a function of model size and then increases after a particular model size.

%\paragraph{ChatGPT}
In 2022, OpenAI introduced a conversational \gls{llm} -- called ChatGPT~\cite{ChatGPT}.
As a model, the first version of ChatGPT was based on \gls{gpt}-3.5 and is an InstructGPT sibling.
\gls{gpt}-3.5 is a \gls{gpt}-3.0 model trained on a training data set that contains text and software code up to the fourth quarter of 2021~\cite{GPT3-5}.
InstructGPT was introduced in \emph{''Training language models to follow instructions with human feedback ''}~\cite{ouyang2022training}, and is a \gls{gpt}-3 model fine tuned with supervised learning in the first step, and in the second step with reinforcement learning from human feedback~\cite{NEURIPS2020_1f89885d}.
ChatGPT is hence a \gls{gpt}-3.5 model fine tuned for conversational interaction with the user.
In other words, the user interacts with the model via sequence of text (the conversation) to accomplish a task.
For example, we can copy and paste a text into ChatGPT's input field and ask it to summarize it.
We can even be more specific, we can say that the summary should be 10 sentences long and be written in a preferred style.
Importantly, if we are unhappy with the result we can ask ChatGPT to refine its own summary without copying and pasting the text it should summarize.
At the moment of this writing, ChatGPT can be used with GPT-4 as the backend \gls{llm}.

% \paragraph{\acrshort{opt}}
% In 2022, Zhang et al. introduced \gls{opt} with the paper \emph{''OPT: Open Pre-trained Transformer Language Models''}
% \footnote{See the \href{https://github.com/facebookresearch/metaseq/tree/main/projects/OPT}{GitHub repository of OPT} for more information}~\cite{zhang2022opt}.
% The main contribution of this paper is that it shares all artifacts created for this paper - including the nine models - for interested researchers, so the impact of \glspl{llm} can be studied by any researcher.
% These models are \gls{gpt}-3 class models in parameter size and performance.

% \paragraph{\acrshort{bloom}}
% Another open \gls{llm} is \gls{bloom} (176 billion parameters), which was developed in the BigScience Workshop~\cite{scao2022bloom}.

%\paragraph{Other large language models}
There are also other large language models. In 2022, Zhang et al. introduced \gls{opt} with the paper \emph{''OPT: Open Pre-trained Transformer Language Models''}
%\footnote{See the \href{https://github.com/facebookresearch/metaseq/tree/main/projects/OPT}{GitHub repository of OPT} for more information}~
\cite{zhang2022opt}.
The main contribution of that paper is that it makes all artifacts including the nine models available for interested researchers.%, so the impact of \glspl{llm} can be studied by any researcher.
These models are \gls{gpt}-3 class models in parameter size and performance.
Another open \gls{llm} is \gls{bloom} (176 billion parameters), which was developed in the BigScience Workshop~\cite{scao2022bloom}.

\subsection{Uptake of large language models}
\label{subsec:business:processes:and:large:language:models}
%
% Write here about promises of large language models.
%
Above, we briefly discuss \glspl{llm}, where we focus particularly on the \gls{gpt} model family as these are the most popular \glspl{llm}, we hypothesise.
It is important to recognize the transition from \gls{gpt}-3 to \gls{gpt}-4, as it brought a massive increase on a variety of benchmarks, particular on academic and professional exams~\cite{OpenAI:GPT4}.
These performance increases in \gls{nlp} tasks are a result of natural language understanding and have, as we argue, massive implications for what can be automated -- the automation frontier.
This frontier is arguably shifted further when natural language understanding is combined with plugin software components.
In fact, at the time of writing, the company behind \gls{gpt} is experimenting with Chat\gls{gpt} plugins.
Among the currently offered plugins are Klarna, %(a web-service for searching and comparing prices from online shops), 
Wolfram, %(Access computation, math, curated knowledge & real-time data through Wolfram|Alpha and Wolfram Language), 
the integration with vector data bases for information retrieval, and an embedded code interpreter for Python~\cite{OpenAI:ChatGPT:plugins}.
This has an impact on \gls{rpa}, and more broadly on business process automation including \glspl{bpms}, and more generally on how work is carried out.

\section{Large language models and the BPM lifecycle}
\label{sec:large:language:models:and:the:bpm:lifecycle}

In this section we identify applications of \gls{llm} within \gls{bpm}.
We systematically explore these applications along the phases of the \gls{bpm} lifecycle, namely identification, discovery, analysis, redesign, implementation, and monitoring~
\cite{dumas2018fundamentals}.
%In particular, we show how \gls{llm} can support \gls{bpm} activities.
In this way, we complement recent efforts to build an overarching inventory of \gls{llm} applications, such as in other fields like %Please note that there are many public repositories that collect \gls{llm} applications for other fields, 
data mining~\footnote{
See for example the \href{https://github.com/openai/openai-cookbook}{OpenAI Cookbook GitHub repository}, which provides code examples for the OpenAI \acrshort{api}}.

\subsection{Identification}
The BPM lifecycle starts from \emph{Identification}. Normally, at this stage there is not much structured process knowledge available in the company, and relevant information has to be extracted from heterogeneous internal documentation. This is exactly where \gls{llm} shine as they can quickly scan and summarize large volumes of text, highlighting important documents or directly outputting required information. 

\paragraph{Identifying processes from documentation}
The idea is to give \gls{llm} all relevant documentation existing in the organization as input. This can include legal documents, job descriptions, advertisements, internal knowledge bases and handbooks. The \gls{llm} is then tasked to identify which processes are taking place in the organization. It can be further instructed to classify the input documents according to processes they describe.
Multimodal \glspl{llm} can improve the results even further as charts, presentations and photos can also directly be used as information sources.

\paragraph{Process selection}
\gls{llm} can be further asked to assess strategic importance of processes based on, e.g. number and types of documents that refer to them as well as extract this information from process descriptions. If given access to information systems supporting the process or other KPIs, \gls{llm} can also assess process health. Finally, assessing feasibility is also theoretically achievable as long as necessary information, e.g. recent technology reports, is given as input as well. Based on these criteria, \gls{llm} can prioritize the processes for further improvement.

\subsection{Discovery}
The second stage of BPM lifecycle is \emph{Process Discovery}. At this stage one or a combination of process discovery methods is selected to produce process models. When one speaks of automated process discovery, one usually means process mining -- a technique of extracting process models and other relevant data out of event logs left by information systems supporting the execution of a process. However, with \gls{llm} also other discovery techniques can benefit from (at least partial) automation.

\paragraph{Process discovery from documentation}
Apart from process mining, documentation analysis is an established process discovery method. In this method, process analyst uses the information found in heterogeneous sources such as internal documentation, job advertisements, handbooks, etc. Searching in these documents might require a lot of time and effort. \gls{llm} are extremely suitable for this task as they can summarize high volumes of text in a concise and structured way. More precisely, they can output process descriptions in desired format (plain text, numbered lists, etc.). One can also specify the level of detail, as to whether the output should include only the activities and events or also resources and additional information. Finally, as some \gls{llm} are also capable of working with structured document formats such as XML, in fact even BPMN models can be produced automatically.

\paragraph{Process discovery from communication logs}
Another information source that can be used in evidence-based discovery is communicaiton logs, i.e. e-mails and chats between process participants: internal employees but also external partners and customers. \gls{llm} can extract patterns from these communication logs, which can be seen as various steps in a process. Then, they can similarly produce process descriptions or models.

\paragraph{Interview chat bot}
Possible applications of \gls{llm} in process discovery can also go beyond evidence-based discovery. Another common discovery method are interviews with domain experts. In these interviews, process analyst asks questions about the process and produces a process model based on several interviews. Typically, several separate interviews with different domain experts are required to produce the first version of process model. Afterwards, additional rounds of interviews are conducted in order to get and incorporate feedback and to perform validation. In the worst case, domain experts might have conflicting perceptions of the process, then resolving such conflicts becomes a very difficult and time-consuming task for both process analyst and domain experts. 

\gls{llm} can solve parts of this problem by providing a chat bot interface for domain experts. In this way, the domain experts answer questions in the chat. This can bring a lot of advantages. First, the domain experts do not have to allocate lengthy time slots for interviews but instead talk with the chat bot at desired pace. Second, the feedback loop gets shorter as \gls{llm} can produce process models directly after or even during the conversation with the domain expert and also do updates to the model, thus validation can happen simultaneously with model creation. Finally, the benefits will only grow if multiple domain experts interact with the chat bot simultaneously (and independently) but the chat bot can use all of this input in the conversations. The latter option is, however, more difficult to implement.

\paragraph{Combined process discovery}
All process discovery methods have their advantages and drawbacks. Often, a combination of these methods is used to achieve best results. However, this combination is limited by the resources that are allocated for process discovery task. Discovery methods presented above give valuable output yet requiring much less resources. Thus, it is possible to apply more of them simultaneously for even better result. The combination of these methods can be used in addition to traditional process mining or "manual" process discovery, which will provide the richest insights. While it could happen that the results of different methods have some inconsistencies that will have to be fixed, also fixing them can be done in (semi-)automated manner.

\paragraph{Process model querying}
As \gls{llm} seem to "understand" process models serialized as XML, they can be used to answer some questions about the model. This can be very useful for quality assurance. First of all, it can be used for checking syntactic quality. While there are tools out there that can do it already, and with much less overhead, it is still convenient to have this feature in \gls{llm} because \gls{llm}, in contrast to other methods, may be able to check other quality aspects as well. For instance, it can also check semantic quality. Indeed, process analyst can give \gls{llm} both interview transcript and a process model as input and \gls{llm} can check both validity and completeness based on this interview. It must be noted, of course, that this will only work under the assumption that the interview transcript has these features of validity and completeness. Another way of checking semantic quality of the model would be via process simulation, e.g. to explicitly ask \gls{llm} whether the given process model could have produced a given execution sequence or to ask \gls{llm} to give possible execution sequences that can be generated by the model.
\gls{llm} are known to be able to simulate Linux shell, for instance, thus they might be also able to simutale a BPMS execution engine as long as enough input is provided.
Finally, \gls{llm} can also (at least so some extent) check pragmatic quality of the models as long as some definition of guidelines, e.g. 7PMG is provided as input as well. It must be also noted that \gls{llm} can not only spot these quality issues but also suggest fixes.

\subsection{Analysis}
The next stage is \emph{Process Analysis}. At this stage, the discovered processes are analyzed to find problems and bottlenecks. While this is a cognitively loaded task, \gls{llm} can be used to help human analysts in some regard.

\paragraph{Issue discovery}
If an issue exists in a process, chances are high somebody has already complained about it. Depending on the company, product, and process it can be the customer, partner or an employee and in can happen on different platforms, including social media, support service or internal communication tools. 
\gls{llm} are good at summarizing large volumes of unstructured text as well as finding patterns, and this capability can be used for this task. It is as easy as just scraping the text from these platforms and giving it as input to the \gls{llm} with a simple prompt like "find all things customers have complained about".

\paragraph{Issue spotting}
After an issue in the process is found, the next step is to spot the part of the process that creates this issue. In some cases, it can be a difficult task, especially in a complex process. 
The idea here is to give \gls{llm} all process models (or models of the relevant process in case it is known that only one process causes the issue and it is known exactly which process) and the spotted problems. The task of \gls{llm} is, by analyzing task names and descriptions to make suggestions which tasks may be responsible for the issue.
In advanced cases, \gls{llm} might be even capable of suggesting some fixes. It might be something as simple as suggesting to automate some manual task that takes too long but it also might be some more complex process redesign suggestion as long as \gls{llm} is given redesign methods as additional input or is trained on redesign methods as well.

\subsection{Redesign}
The fourth phase of the BPM lifecycle is \emph{Process Redesign}. In this stage, process improvement suggestions are developed based on discovered issues and general process improvement methods. These suggestions are evaluated, and a to-be process model is developed at the end of this stage.

\paragraph{Business process improvement}
An obvious yet very promising use case is t just ask \gls{llm} to redesign the process. As already mentioned, simple issues arising from just one activity can be fixed by the \gls{llm}. However, it does not stop there and is theoretically only depending on the quality of the input given to the \gls{llm}. Indeed, if it is given exhaustive information about the process (detailed process model as well as description of the process or tasks) as well as detailed description of some redesign method (or it is trained on some redesign methods), redesigning the business process is as simple as just telling the \gls{llm} to apply the method on the process. This can, however, be improved even further. First, the description of the issues discovered in the previous phase can be given as additional input to guide process redesign to fix those first. Second, \gls{llm} can be instructed to apply different redesign methods and to give separate lists of suggestions given by each of them so the analyst can then select the best options. Moreover, \gls{llm} itself can be asked to choose the best suggestions and motivate its choice.
It must be noted, of course, that this will only work if sufficient input is given. For instance, for inward-looking redesign methods, the methods themselves as long as detailed process information is required. For outward-looking methods, in addition to that, there should be enough outside information and/or a way for \gls{llm} to properly communicate with the outside world.

% \paragraph{Business process simulation}
% \todo{todo}

\subsection{Implementation}
The next phase of the BPM lifecycle is \emph{Process Implementation}. It covers organizational and technical changes required to change the way of working of process participants as well as IT support for the to-be process.

\paragraph{BPMN model explanations with plain text}
As mentioned, \gls{llm} can work with BPMN models serialized in XML. We have already discussed how \gls{llm} can manipulate process models in order to increase quality as well as suggest or incorporate redesign ideas. To close the circle, \gls{llm} can produce textual explanations of BPMN models. What is more interesting, one can control the level of detail as well. So, depending on the target audience, \gls{llm} can produce textual overview  but also detailed descriptions of the models. It can transform it into requirements for software developers if enough details are contained in the BPMN model itself.

\paragraph{BPMN model chatbot}
Building on top of the previous use case, model description can be also tailored to every specific user. This way, given a model or -- better -- model repository with additional documentation, \gls{llm} can prepare specific descriptions for, e.g. process owner but also for individual participants for which all specific tasks they are responsible for are also described and explained in detail. Furthermore, in this use case one can add interaction between the user and \gls{llm}. This way, user may ask clarifications for parts he did not understand or generally ask for more details as long as some guidance is required.

\paragraph{Process orchestrator}
\glspl{llm} can be accessed via APIs and at the same time can access APIs themselves, opening a huge variety of opportunities. While the former means it can be used for automated tasks and be called by the orchestrator, the latter means that it could theoretically be an orchestrator itself: given executable process model and additional constraints as context as well as the required instance data as input, it can theoretically execute a process by calling other APIs and assigning tasks in a more flexible way than a traditional orchestrator.
% \subsection{Execution}

\subsection{Monitoring}
The last phase of the BPM lifecycle is \emph{Process Monitoring}. At this stage, already implemented processes are executed, and their performance is monitored. The observations collected in this phase are used for operational management as well as serve as input for further iterations of the lifecycle.

\paragraph{Process dashboard chatbot}
Dashboards are a powerful tool that provides overview of the most important KPIs of a process on a single screen. However, the ultimate goal of them is to tell the viewer whether the status of the process is good or not, and the numbers and colors are mostly used as an intermediary medium. \gls{llm} can take away this intermediate step and allow the user to directly know the status of the processes. 

% \subsection{Adaption and evolution}

\section{Research directions}
\label{sec:research:directions}

In this section we propose the research directions.
We categorize the research directions into three groups. 
The first group studies the use of \gls{llm}, and their applications, in practice.
This includes the use within \gls{bpm} projects in companies or as part of an \gls{is} (\Cref{subsec:llm:practice}), the development of usage guidelines for practitioners and researchers (\Cref{subsec:guidelines}), and also the derivation of \gls{bpm} tasks (\Cref{subsec:tasks:specific:to:bpm}) and their corresponding data sets (\Cref{subsec:data:sets:bpm}).
The second group studies how \gls{llm} can be combined with existing \gls{bpm} tools, and more generally \gls{bpm} technologies, to increase user experience (\Cref{subsec:llm:bpm:combination}).
Crucially, this group draws from findings in the first group.
The third and final group develops large language models specifically for business process management, so these models can understand the context and language of business processes and support various tasks, such as process discovery, monitoring, analysis, and optimization (\Cref{subsec:develop:llm:for:bpm}).
Again, this group builds upon the findings of the first group.

\subsection{The use of large language models in \acrshort{bpm} practice}
\label{subsec:llm:practice}
% We created a list (section 3), the quetion is now:
% -  is this list complete?
% - what parts of the list make sense?
% - what parts bring the most value?
% - can we already implement them with existing models?
% - do we always need the biggest and best models? (most probably not)
% - how do LLM change BPM projects as socio-technical systems?
The first research direction studies the use of \gls{llm} in practice.
One major question to answer is for which tasks \gls{llm} can be used.
In \Cref{sec:large:language:models:and:the:bpm:lifecycle}, we present a list of tasks for which \gls{llm} can be used.
However, this list might not be complete, in addition some of the tasks might turn out to be of little use.
Tied to this is the question what tasks will bring, and ultimately bring the most value for an organization.
The next big question is the relation between a task and the model properties needed to achieve a pre-defined value.
One question here is which tasks can be achieved with already existing models.
Another question to study is whether we always need the largest, and hence most accurate model, for each task.
We hypothesize that this might not be the case.
Finally, and most importantly, the next big question to answer is how \gls{llm} will change how work is carried out within \gls{bpm} projects, and within processes that are actively managed.
We for example hypothesize that conversational \glspl{llm} might take the spot of the duck in the famous \emph{duck approach} \footnote{
    \href{https://en.wikipedia.org/wiki/Rubber_duck_debugging}{Rubber duck debugging}
}.
This question is a socio-technical systems question, and we hence strongly believe that the \gls{bpm} community, and the information systems community more broadly, is especially well equipped to contribute to this question.

\subsection{Usage guidelines for researchers and practitioners}
\label{subsec:guidelines}
%
% Outline
%
% model selection: given a task, how to select the best model, which characteristics to look at
% patterns repository: based on possible tasks in each lifecycle phase and specific context of the organization, propose parts that are the most relevant to implement/ have to be implemented first.
% Example: process monitoring - based on real process executions and organizational goals, prescriptive process mining might be urgent to implement in company ABC, but not necessarily in other case
% \paragraph{Prompt engineering}
% Usage guidelines also include systematically creating and collecting best practices for prompts.

The second research direction builds usage guidelines for 
\gls{bpm} researchers and practitioners.
One question such guidelines have to answer is given an organizational context, the lifecycle phase, and the process context of a task, suggest a \gls{llm} to achieve an expected value.
In addition, such guidelines systematically collect best practices for creating prompts.
For example, for the \gls{bpm} lifecycle phase process implementation, and monitoring and controlling, a company might consider using a \gls{llm} within a managed process.
Let us assume this company is a bank and wants to automate the task of replying to customer inquiries with \gls{llm}.
Then this guideline proposes for the process implementation a specific \gls{llm}, with the number of parameters it has, gives examples on how to create a prompt template, fill the template with customer background information, and finally on how to integrate the customer inquiry within the prompt template.
For process monitoring and controlling, the guidelines might propose a different model for analyzing different inquiry clusters as the lifecycle phase context is different.
As an example, consider here that the \gls{llm} first categorizes each inquiry into a positive and negative sentiment, and then lists for both the top five inquiry reasons.
This research direction builds upon the first research direction, as first research direction, among others, determines the tasks for which \gls{llm} can be used in principle.

\subsection{Creation, release, and maintenance of task variants specific to \acrshort{bpm}}
\label{subsec:tasks:specific:to:bpm}
This research direction builds and maintains two different task lists.
The first list maps general \gls{nlp} tasks to tasks within \gls{bpm}.
As an example, consider the general \gls{nlp} task of text summarizing.
Within \gls{bpm}, text summarizing can relate to summarizing a set of process descriptions or task descriptions.
We can think of this list as a one to many mapping between \gls{nlp} tasks on the one hand, and \gls{bpm} tasks on the other.
The second list enumerates tasks that are unique to \gls{bpm}.
% Example?
This research direction uses the findings from the directions presented in \Cref{subsec:llm:practice} and \Cref{subsec:guidelines}.

\subsection{Creation, release, and maintenance of data sets and benchmarks}
\label{subsec:data:sets:bpm}
Public data sets and benchmarks are crucial for the progress of \gls{llm} in research as they allow researchers to measure progress.
In addition, they are also important for practitioners as they define data set properties (such as meta-information) they are likely to need themselves when they fine tune a model.
As a result, data sets and benchmarks need to be properly aligned with the automation needs of \gls{bpm}.
Blagec et al. argue similarly as we, but for the clinical profession~\cite{blagec2022benchmark}.
In their study, they analyzed 450 \gls{nlp} data sets and found that \emph{''AI benchmarks of direct clinical relevance are scarce and fail to cover most work activities that clinicians want to see addressed''}.
A research direction for the \gls{bpm} community is hence to do the same for \gls{bpm}.
One question worth studying is whether existing \gls{nlp} data sets and benchmarks are of relevance to \gls{bpm}, for example, if they cover the activities of \gls{bpm} researchers and practitioners.
This research direction builds upon the research direction in \Cref{subsec:tasks:specific:to:bpm}.

\subsection{\acrshort{llm} and \acrshort{bpm} artifacts}
\label{subsec:llm:bpm:combination}
% which artifacts do we actually create in BPM and how can LLM use them?
% multimodal
% context engineering: how should the context (e.g. of a BPMN model) should look like so that LLM can produce meaningful output. Specific case of prompt engineering
% Multimodal prompt engineering for BPM
This research direction studies the interplay of \gls{llm}, \gls{bpm} artifacts, and \gls{bpm} tasks.
The goal is to understand which artifacts are necessary for \gls{llm}, and their multimodal successors, to create useful outputs.
It can hence be understood as a special case of prompt engineering, which we might call multimodal prompt engineering for \gls{bpm}.
This is an important research direction as the output quality of a \gls{llm} depends heavily on the context quality and quantity it is given.
In other words, the more context, and the higher the quality of each context, the higher the output quality of the \gls{llm}.
For this reason, we believe that it should be considered  its own research direction.
As an example, consider again the customer inquiry process from above.
In this case, we can imagine that the context of the \gls{llm} depends on the inquiry.
In one case, the customer might include an image in the inquiry.
Or think of the redesign phase of the inquiry process.
During this phase, artifacts are created, for example drawings of processes on a board, comments to these processes in a word processor, and remarks on data availability and access in an audio file.
This information might be useful when we ask -- a possibly different -- \gls{llm} why a customer inquiry on current special offers cannot yet be answered.
The reason here might be that a central system which stores special offers does not yet exist.
This research direction builds upon the directions presented in \Cref{subsec:llm:practice} and \Cref{subsec:guidelines}.

\subsection{Development and release of \acrlong{llm}s for \acrlong{bpm}}
\label{subsec:develop:llm:for:bpm}
% This research area could lead to the creation of dedicated models that can better handle the specific challenges and requirements of \gls{bpm}, compared to general-purpose language models.
% Please note here that we do not mean specialized models in the sense of exclusive for, but rather general-purpose language models that are also trained on the \gls{bpm} domain.
This research directions studies how \gls{llm} are build for \gls{bpm} tasks, all previously discussed research direction are the foundation for this direction.
The goal of the research direction is to build \glspl{llm}  that are attuned to the specific challenges and requirements of \gls{bpm}, compared to general-purpose language models.
This includes specialized models in the sense of exclusive for, and also general-purpose language models that are fine-tuned on the \gls{bpm} domain.
An important aspect of this direction is to open source the created \gls{llm}, as is done for \gls{opt}~\cite{zhang2022opt}.
This is important for researchers can use this model in their studies, and practice as companies can use these models free of charge for their use cases.

\section{Discussion}
\label{sec:discussion}
%
% Version 2
%
In this section we discuss the challenges of \gls{llm}, the power of combination and inflated expectations, and end with an outlook and future work.

\paragraph{Challenges}
The use of \gls{llm} entails opportunities and challenges. 
For example, they can help to understand difficult research, but they also carry over deficiencies (including factual errors) in the training data set to the texts they generate~\cite{van2022language}.
In a systematic study of these errors, Borji analyzes errors of ChatGPT and categorizes them -- the author further outlines and discusses the risks, limitations and societal implication of such models\footnote{See the \href{https://github.com/giuven95/chatgpt-failures}{ChatGPT failure archive (GitHub)} for an up-to-date list}~\cite{https://doi.org/10.48550/arxiv.2302.03494}.
The failure categories identified by the author include reasoning, factual, math, and coding. 
A similar deficiencies study was done in~\cite{10.1145/3442188.3445922}, but these authors focus on \gls{llm} in general.
A news feature in Nature discusses these and the risks of using \gls{llm}~\cite{Hutson2021-bl}.
One consequence for education might be that essays as an assignment should be re-considered~\cite{stokel2022ai}.

\paragraph{The power of combination and managing expectations}
The major innovation of Chat\gls{gpt} was not the introduction of a new technology, but the combination of already existing ones and an easy to use user-interface~\cite{OpenAI:Combination}.
This effect of combination extends beyond \gls{llm}, \gls{nlp}, or \gls{ml} innovations.
For example, OpenAI is currently experimenting with integrating ChatGPT with software plugins, which might even in the short run lead to a software marketplace for their platform\footnote{\href{https://openai.com/blog/chatgpt-plugins}{https://openai.com/blog/chatgpt-plugins}}.
For this reason, we suggest and advocate in our research directions above to study and build these combinations with \emph{existing} \gls{bpm} technologies, instead of solely focusing on developing new ones.
In this paper, we have so far made the case for the opportunities \gls{llm} realize, shortly discussed their shortcomings, and pointed out how important it is to combine technologies within a field, and across field boundaries.
However, we also stress here how important it is to manage, maybe even overshooting, expectations driven by this very recent developments.
For example, the speculation about the possible capabilities on the successor of \gls{gpt}-3 were driven up by the hype to a point where \emph{''people are begging to be dissapointed''}~\cite{OpenAI:Combination}.

%
% Included as ref.
%
 %\footnote{This was noted by Sam Altman - the CEO of OpenAI - in a conversation (\href{https://www.youtube.com/watch?v=ebjkD1Om4uw}{https://www.youtube.com/watch?v=ebjkD1Om4uw})}.

\paragraph{Outlook and future work}
\gls{llm} are used, and will be used in commercial products with huge amounts of users.
We speculate that this will have an effect on research, as funding agencies might increase the amount of grants for this research field.
An ever increasing user base that interacts with \gls{llm} (directly or indirectly) is therefore, in our view, inevitable.
For future work, we plan to work on developing research directions that are beyond the scope of this paper.
We expect that \gls{llm} will have an effect on how work is carried out (see \Cref{subsec:business:processes:and:large:language:models} and \Cref{subsec:llm:practice}).
But this may have far greater impacts than what we cover here, for example on the \acrshort{bpm} capabilities, which are strategy, governance, information technology,
people, and culture~\cite{rosemann2015six}.

\section{Conclusion}
\label{sec:conclusion}

In this paper we present six research directions for studying and building \glspl{llm} for BPM.
We use the BPM lifecycle to propose applications of LLM to showcase the impact of these models.

%\subsubsection{Acknowledgements} 
%This work received funding from the Teaming.AI project in the European Union's Horizon 2020 research and innovation program under grant agreement No 95740. The research by Jan Mendling was supported by the Einstein Foundation Berlin under grant EPP-2019-524 and by the German Federal Ministry of Education and Research under grant 16DII133.

%
% ---- Bibliography ----
%
% BibTeX users should specify bibliography style 'splncs04'.
% References will then be sorted and formatted in the correct style.
%
\bibliographystyle{splncs04}
\bibliography{bib/references.bib}

\end{document}